\documentclass[apj]{emulateapj}
\usepackage{amsmath}
\usepackage{amssymb}
\usepackage{graphicx}
\usepackage{bm}

\begin{document}

\title{SCALING OF THE GROWTH RATE OF MAGNETIC ISLANDS IN THE HELIOSHEATH}

\author{K. M. Schoeffler} \author{J. F. Drake} \author{M. Swisdak} 
\affil{Institute for Research in Electronics and Applied
Physics, University of Maryland, College Park, MD 20742-3511, USA}

\date{\today}

\begin{abstract}

Current sheets thinner than the ion inertial length are unstable to the tearing
instability and will develop magnetic islands that grow due to magnetic
reconnection. We investigate whether the growth of magnetic islands in a
current sheet can continue indefinitely, or in the case of the heliosheath until
reaching a neighboring current sheet, and at what rate the islands grow. We
investigate the development and growth of magnetic islands using a
particle-in-cell code, starting from particle noise. Performing a scaling of
the growth of magnetic islands versus the system size, we find that the growth
rate is independent of the system size up to the largest simulation we were
able to complete. The islands are able to continue growing as long as they
merge with each other and maintain a high aspect ratio. Otherwise there is not
enough magnetic tension to sustain reconnection. When applied to the sectored
magnetic fields in the heliosheath, we show that the islands can continue
growing until they reach the sector width and do so in much less time than it
takes for the islands to convect through the heliosheath.

\end{abstract}

\maketitle

At low latitude the solar wind is divided by the heliospheric current sheet
into sectors of oppositely directed magnetic fields. The thickness of the
current sheet, $\lambda$, is around $10,000 \text{km}$ \citep{Smith01} at $1
\text{AU}$, and the separation between each sector, or the sector width, is
around $1 \text{AU}$. The sector width remains nearly constant, although
increasingly variable, out to the termination shock (TS) at $\sim 90
\text{AU}$, where the supersonic solar wind abruptly slows
\citep{Behannon89,Burlaga07}. For the essentially collisionless environment of
the solar wind $\lambda$ controls whether collisionless reconnection onsets.
For $\lambda$ greater than the ion inertial scale, $d_i = c/\omega_{\text{pi}}$, where
$c$ is the speed of light and $\omega_{\text{pi}}$ is the ion plasma frequency,
reconnection via the collisionless tearing instability does not take place,
while for $\lambda < d_i$ it does \citep{Cassak05,Yamada07}. Just upstream of
the TS where the ion density is $\sim 0.001\text{cm}^{-3}$, $d_i \sim 7200 \text{km}$,
which remains smaller than $\lambda$ based on measurements at $1 \text{AU}$.
Thus, the absence of significant reconnection of the sector field upstream of
the TS is consistent with models, although a definitive study of $\lambda$
upstream of the TS remains to be carried out. $\lambda$ downstream of the TS is
predicted to be $2500 \text{km}$ based on measurements at $1 \text{AU}$ and
the shock compression, while $d_i$ is $4200 \text{km}$ based on ion density
measurements of about $0.003 \text{cm}^{-3}$ \citep{Richardson08}. Thus,
downstream of the TS the current sheets should begin breaking up into magnetic
islands.

The growth of islands in the finite plasma $\beta$ (the ratio of the plasma
pressure to the magnetic pressure) heliosheath has been investigated in
particle-in-cell simulations \citep{Drake10,Opher11,Schoeffler11}. The
predicted island length in the initial phase of reconnection is around $190
d_i$, much smaller than the sector width, $8900 d_i$, so it is uncertain
whether islands will grow to the full sector width. Due to computational
limitations it is not possible to realistically simulate the disparate ion
inertial and sector scales. In this work, we perform a scaling study of the
growth of islands, in which the inter-current sheet separation (equivalent to
the sector width) varies, in order to understand what happens in the real
system. Even our largest simulations have sector widths that are much smaller
than in the actual sectored heliosheath.

The current sheets separating the sectored regions begin to form islands after
crossing the TS. As the islands grow, the current sheets are convected toward
the heliopause. The plasma flows outward at around $80 \text{km s}^{-1}$ and
steadily decreases in speed for $20 \text{AU}$, at which point the radial flow
remains close to zero \citep{Krimigis11}. The important question which we seek
to answer is whether the islands are able to expand to the sector width before
the current sheet reaches the heliopause. If the islands expand to the sector
width, there would be no more laminar field that can shield cosmic rays, and
cosmic rays could percolate through the system consisting solely of islands. In
addition, the full expansion of the islands would imply that acceleration
mechanisms due to the growth of islands may play a significant role.

We simulate the growth of magnetic islands using a particle-in-cell code, p3d.
The initial conditions consist of two oppositely directed current sheets in
Harris equilibrium\citep{Harris62}, with a superimposed background density.
The initial magnetic fields are in the $\mathbf{\hat{x}}$ direction, which
corresponds to the azimuthal direction in the heliosheath. The current flows in
the $\mathbf{\hat{z}}$ direction, which corresponds to the north south
direction. The $\mathbf{\hat{y}}$ direction corresponds to the radial direction
of the heliosheath. In the heliosheath the islands are predicted to be highly
elongated due to the development of pressure anisotropy, and this elongation is
dependent on both the ion-to-electron mass ratio and the electron temperature
\citep{Schoeffler11}. Typical simulations use a reduced mass ratio in order to
reduce computational expenses, which produces much shorter islands than
expected for the real system. We therefore use an enhanced temperature of the
electrons in the background in order to form more realistic elongated islands.
In the heliosheath nearly all of the pressure comes from the ions due to the
pick-up ion population. In our simulation half of the pressure is in the ions
and half is in the electrons. The $\beta$ of the plasma in the asymptotic field
is $3$, which corresponds to a temperature in the heliosheath of $650,000
\text{K}$, based on a density, $n = 0.003 \text{cm}^{-3}$, and a magnetic
field, $B = 0.15 \text{nT}$. The ratio of the proton to electron mass in this
simulation is $25$. We have run simulations that imply the reconnection rate is
not sensitive to the value of the mass ratio, so our prediction should scale to
the realistic mass ratio. The background temperature is $15$ times the Harris
sheet temperature of $0.25 m_ic_A^2$, where $m_i$ is the ion mass and $c_A$ is
the Alfv\'en speed, based on the asymptotic magnetic field, $B_0$. The ratio of
the speed of light to the Alfv\'en speed is $25$.  Each simulation has a grid
scale resolution of $\Delta_x = \Delta_y = 0.05d_i$ and a time resolution of
$dt = 0.004\Omega_{\text{ci}}^{-1}$, where $\Omega_{\text{ci}}$ is the ion cyclotron
frequency. The half thickness of the current sheet is set to $w_0 = 0.5 d_i$ so
that collisionless reconnection can begin from particle noise. We simulate a two-dimensional system. In three-dimensional systems islands form at different $z$
locations and grow along $z$, eventually stagnating likely due to interactions
with other islands \citep{Shay03,Schreier10}.  The significance of this effect
in the heliosheath is unknown.

\begin{figure}
  \noindent\includegraphics[width=3.0in]{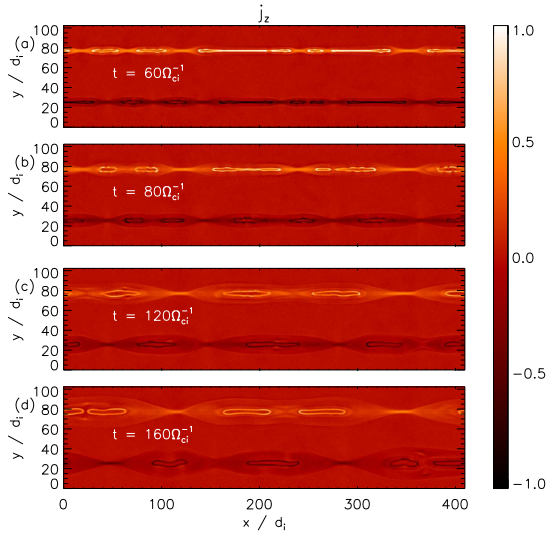}
  \caption{\label{merging}
  Out-of-plane current density, $j_z$, for $t = $ (a) $60\Omega_{\text{ci}}^{-1}$, (b)
  $80\Omega_{\text{ci}}^{-1}$, (c) $120\Omega_{\text{ci}}^{-1}$, and (d)
  $160\Omega_{\text{ci}}^{-1}$ For better contrast, all points with $|j_z|>1$ are
  assigned the colors shown for $1$ or $-1$.
}
\end{figure}

The time evolution of the largest of these simulations, with dimensions of
$409.6 d_i \times 102.4 d_i$, can be seen in Figure \,\ref{merging}. By $t =
60\Omega_{\text{ci}}^{-1}$ the current sheet has broken into elongated magnetic
islands as predicted in \cite{Schoeffler11} (Figure \,\ref{merging}(a)). The
length of the islands is smaller than the separation between the two current
sheets, so it is expected that they could not grow to the sector width since
circular islands do not have tension to drive reconnection. However, as can be
seen in the subsequent times (Figure \,\ref{merging}(b)-(d)), the islands
on a given current sheet begin to merge. Merging lengthens the islands
which enables further growth until they approach the
neighboring current sheet.

\begin{figure}
  \noindent\includegraphics[width=3.0in]{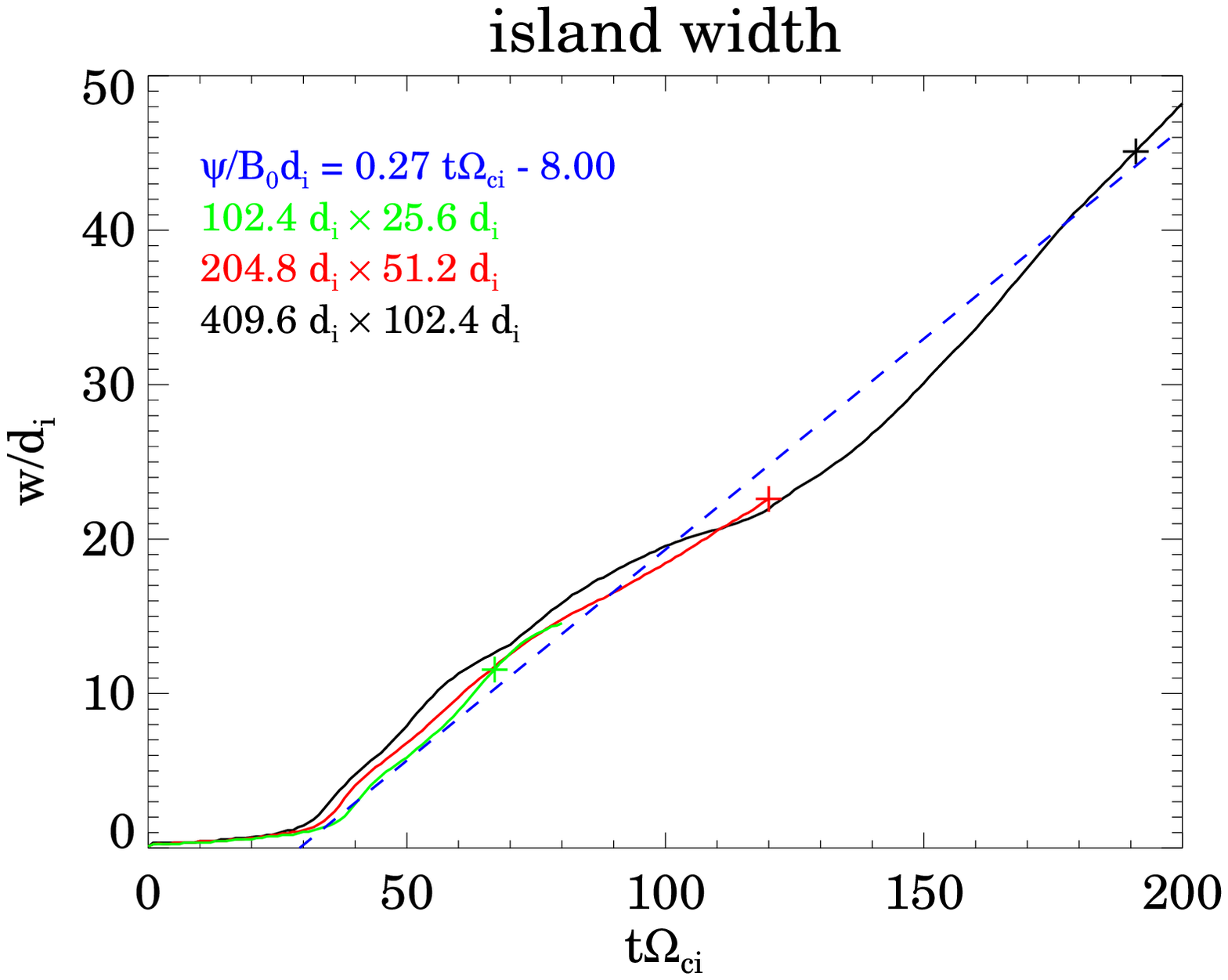}
  \caption{\label{islandwidth}
  Width of the largest island in the lower of the two current sheets vs. time
  for simulations with dimensions of $102.4 d_i \times 25.6 d_i$ (green), $204.8
  d_i \times 51.2 d_i$ (red), and $409.6 d_i \times 102.4 d_i$ (black). The
  plus signs denote the point where the island width reaches $44\%$ of the
  system size and represent the time when the island begins to interact with the other
  current sheet. The blue dashed line is a line of best fit between these three
  points. The slope of this curve is the rate of growth at which the
  island expands, $0.12 c_{\text{Ab}}$.
}
\end{figure}

The islands found in the heliosheath, which are predicted to be much shorter
than the current sheet separation, should in principle grow to the sector
width as long as the islands are able to continue merging. The aspect ratio of
our simulation box is $4$, which was sufficient for islands to continue merging
up to the time that they begin interacting with the neighboring current sheet.
We also carried out a simulation with aspect ratio $2$ where the islands
stopped merging when their length approached the system size. The islands
saturated and stopped growing when there was no longer enough magnetic tension
to maintain reconnection. The size of the heliosheath in the azimuthal
direction greatly exceeds the sector width. Islands should therefore grow to
the sector width. The next question is how long it takes for these islands to
reach the sector width. In order to establish this time we carry out a scaling
of the time required to grow to the sector width for system sizes $102.4 d_i
\times 25.6 d_i$, $204.8 d_i \times 51.2 d_i$, and $409.6 d_i \times 102.4
d_i$.

To determine the size of the island we find the minimum of the flux function ,
$\psi$ where the magnetic field, $\mathbf{B} = \mathbf{\hat{z}} \times
\mathbf{\nabla} \psi\left(x,y\right) + B_z\left(x,y\right) \mathbf{\hat{z}}$,
along the center of the lower initial current sheet at a particular time. This
minimum corresponds to the most developed $x$-point. The upper current sheet has
$x$-points at the maxima. The line of constant flux that crosses an $x$-point is
known as the separatrix. The distance between the maximum and minimum
$y$-locations of the separatrix is defined as the island width, $w$. Figure
\,\ref{islandwidth} shows the island width versus time for each of the
simulations. We use the time when the island reaches $44\%$ of the size of the
box as a measure for when the island reaches the neighboring current sheet. At
$50\%$ the island begins to be affected by the presence of the neighboring
current sheet. The best-fit line connecting the times when the islands reach
the neighboring current sheet fits very well with the island width versus time
for all the simulations.

The results of Figure \,\ref{islandwidth} suggest that the islands grow at a
nearly constant rate that is independent of the system size. Keeping in mind
that this number is based on only three data points, using the slope of the
best fit curve we can obtain an estimate for the rate of growth of the island
of around $0.12 c_{Ab}$, where $c_{Ab}$ is the Alfv\'en speed based on the
background density, which is distinct from the normalization of the code, $c_A$,
which uses the peak density of the Harris equilibrium. If we extrapolate the
growth rate to a very large system we can predict a time for the islands to
reach the sector width. Based on a magnetic field, $B$, of $0.15 \text{nT}$ and
density, $n$, of $0.003 \text{cm}^{-3}$ the Alfv\'en speed just downstream of
the TS is $60 \text{km s}^{-1}$. Using this speed for $c_{Ab}$, and the sector
width, $W=0.25 \text{AU}$, we obtain a growth time, $t_g$, of about $60$ days,
much less than the plasma convection time across the heliosheath.

\begin{equation}
t_g = 60\text{ days}
\left(\frac{W}{0.25 \text{ AU}}\right)
\left(\frac{n}{0.003 \text{ cm}^{-3}}\right)^{1/2}
\left(\frac{B}{0.15 \text{ nT}}\right)^{-1}
\end{equation}
Assuming the radial velocity of the solar
wind inside the heliosheath is $70 \text{km s}^{-1}$ during island growth, this time
corresponds to a distance of $2.5 \text{AU}$ past the TS.

\begin{figure}
  \noindent\includegraphics[width=3.0in]{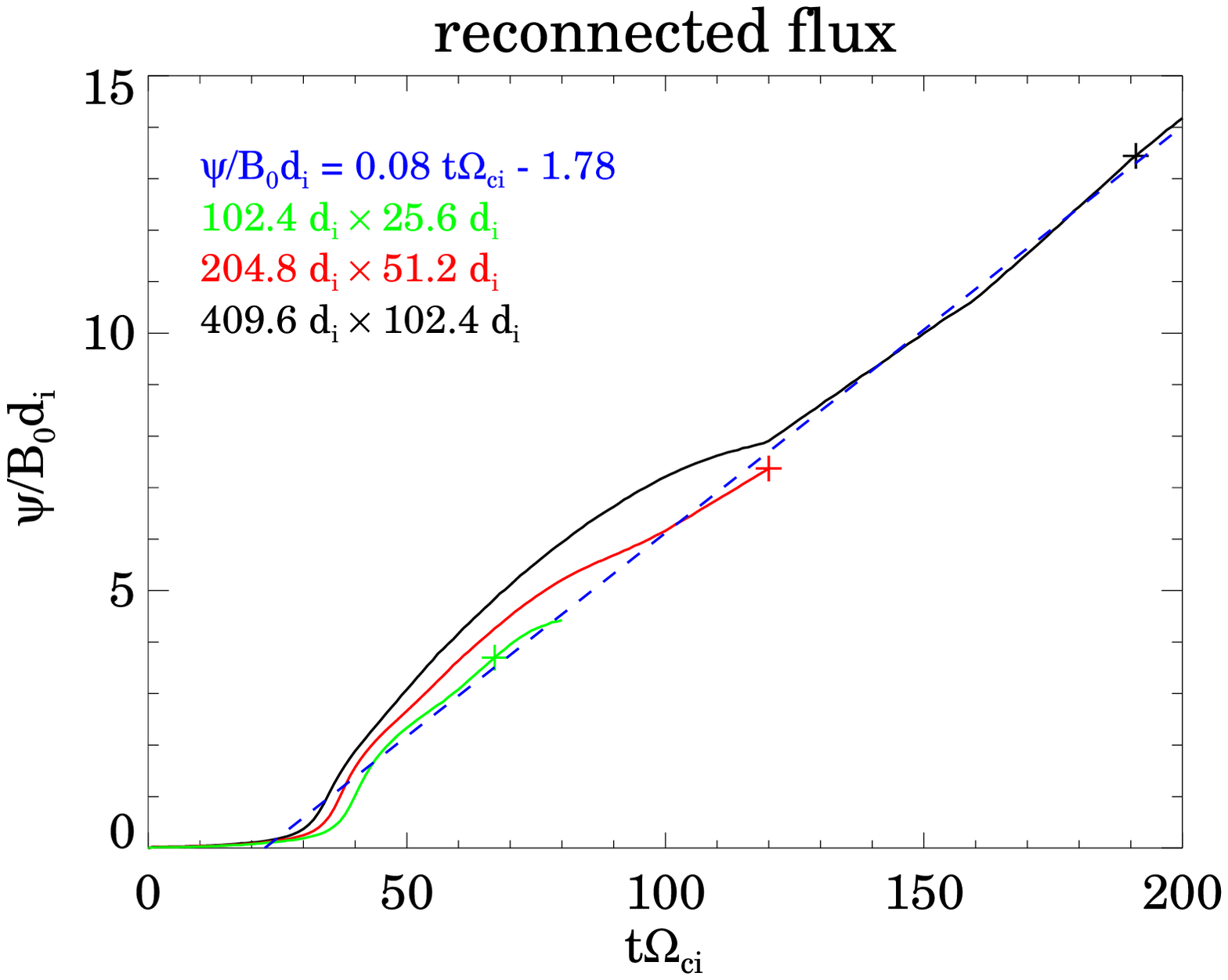}
  \caption{\label{recflux}
  Reconnected flux in the lower of the two current sheets vs. time
  for simulations with dimensions of $102.4 d_i \times 25.6 d_i$ (green), $204.8
  d_i \times 51.2 d_i$ (red), and $409.6 d_i \times 102.4 d_i$ (black). The plus
  signs denote the point where the island width reaches $44\%$ of the system
  size and represent the time when the island begins to interact with the other
  current sheet. The blue dashed line is a line of best fit between these three
  points. The slope of this curve is the reconnection rate $0.079 B_0 c_A$.
} 
\end{figure}

In addition to the rate of change in the island width being nearly constant,
we find that the reconnection rate is also independent of system size. In
Figure \,\ref{recflux} we look at a similar plot to Figure \,\ref{islandwidth},
but of the reconnected flux. To find the reconnected flux we take the
difference between the maximum and minimum of the flux function along the
center of the initial current sheet. Based on the slope of the best fit curve,
the reconnection rate was $0.079$. Previous scalings of reconnection rate
versus system size for conditions with lower plasma $\beta$, relevant to the
Earth's magnetosphere and the $1 \text{AU}$ solar wind, have shown comparable
rates \citep{Shay99, Shay07}. However it is important to note that the
reconnection rates shown in \cite{Shay99, Shay07} are associated with a steady
state reconnection, as opposed to the rate shown here which includes the
complicated dynamics of the merging process as well.

In the range of sizes simulated the rate of flux reconnection and the rate of
island growth are nearly constant once reconnection begins. The growth rates are
independent of the system size. It is reasonable to conclude that in a larger
system these trends would continue. The merging of magnetic islands
allows the islands to maintain a high aspect ratio, which maintains the
magnetic tension necessary to drive reconnection. The steady reconnection rate
allows for a constant rate of island growth, resulting in islands with a width
that scales like the current sheet separation. These islands would be fully
grown long before reaching the heliopause. The growth of these islands in the
heliosheath is vital for the generation of anomalous cosmic rays (ACRs) by
Fermi acceleration in islands \citep{Drake10}. Since these islands are expected
to be present in the sectored region, and the flux of ACRs is greatly reduced
outside of the sectored region \citep{Opher11}, both observations and models
suggest that the sectored heliosheath has broken into magnetic islands.

\smallskip

Computations were performed at the National Energy Research Scientific
Computing Center. This work has been supported by NSF grant ATM-0903964.


\end{document}